\RequirePackage{amsmath}
\documentclass[runningheads]{llncs}
\usepackage[T1]{fontenc}
\usepackage{graphicx}
\usepackage[utf8]{inputenc} 
\usepackage[T1]{fontenc}    
\usepackage{url}            
\usepackage{fancyhdr}
\usepackage{booktabs}       
\usepackage{amsfonts}       
\usepackage{nicefrac}       
\usepackage{microtype}      
\usepackage{lipsum}
\usepackage{graphicx}       
\graphicspath{{media/}}     
\usepackage[colorlinks=false, pdfborder={0 0 1}, linkbordercolor={0 1 0}]{hyperref}

\usepackage[nonumberlist,nogroupskip,xindy]{glossaries}
\usepackage[english]{babel}
\usepackage{amssymb}
\usepackage{float}
\usepackage{tabularray}
\usepackage{lineno}
\usepackage{graphicx}
\usepackage[font=small,labelfont=bf]{caption}
\usepackage{appendix}
\usepackage{tabu}
\usepackage{nomencl}
\usepackage{wrapfig}
\usepackage{fancyhdr}
\usepackage{lmodern}
\usepackage{multicol}
\usepackage{diagbox}
\usepackage{colortbl}
\usepackage{booktabs}
\usepackage{hhline}
\usepackage{tabularx}
\usepackage[margin=1in]{geometry}
\usepackage{parskip}
\usepackage{setspace}
\usepackage{booktabs}
\usepackage{multirow}
\usepackage{graphicx} 
\usepackage{lineno}
\usepackage{multirow}
\usepackage[table]{xcolor}
\usepackage{tikz}
\usepackage{hyperref}       
\usepackage{silence}
\usepackage{enumitem}
\usepackage{calc,cleveref,crossreftools}

\begin{document}

\title{Generative Style Transfer for MRI Image Segmentation: A Case of Glioma Segmentation in Sub-Saharan Africa}

\titlerunning{Generative Style Transfer for SSA Glioma Segmentation}

\author{Rancy Chepchirchir\inst{1,2}$^{\dagger}$\and
Jill Sunday\inst{3}$^\dagger$ \and
Raymond Confidence\inst{4,5} \and
Dong Zhang\inst{6} \and
Talha Chaudhry\inst{7} \and
Udunna C. Anazodo\inst{4,5,8,9} \and
Kendi Muchungi\inst{10} \and
Yujing Zou\inst{11}$^\dagger$}

\authorrunning{R. Chepchirchir and Sunday et al.}

\institute{Institute of Mathematical Science, Strathmore University, Kenya \and
Faculty of Science and Engineering, University of Hull, England \newline
\email{r.chepchirchir-2023@hull.ac.uk} \and
Department of Medical Engineering, Technical University of Mombasa, Kenya \newline
\email{jillselesa35@gmail.com} \and
Medical Artificial Intelligence Lab, Lagos, Nigeria \newline
\email{raymondconfidence@gmail.com} \and
Lawson Health Research Institute, London, Ontario, Canada \and
Department of Electrical and Computer Engineering, University of British Columbia, Vancouver, BC, Canada \newline
\email{donzhang@ece.ubc.ca} \and
University of Nairobi, Nairobi, Kenya \newline
\email{talhahchaudhry99@gmail.com} \and
Department of Clinical and Radiation Oncology, University of Cape Town, Cape Town, South Africa \and
Montreal Neurological Institute, McGill University, Montreal, Canada \newline
\email{udunna.anazodo@mcgill.ca} \and
Brain Mind Institute, The Aga Khan University, Nairobi, Kenya \newline
\email{kendi.muchungi@aku.edu} \and
Medical Physics Unit, McGill University, Montreal, Canada \newline
\email{yujing.zou@mail.mcgill.ca}}

\maketitle

\footnotetext{$^\dagger$ Corresponding authors}

\begin{abstract}
In Sub-Saharan Africa (SSA), the utilization of lower-quality Magnetic Resonance Imaging (MRI) technology raises questions about the applicability of machine learning (ML) methods for clinical tasks. 
This study aims to provide a robust deep learning-based brain tumor segmentation (BraTS) method tailored for the SSA population using a threefold approach. Firstly, the impact of domain shift from the SSA training data on model efficacy was examined, revealing no significant effect. Secondly, a comparative analysis of 3D and 2D full-resolution models using the nnU-Net framework indicates similar performance of both the models trained for 300 epochs achieving a five-fold cross-validation score of 0.93. Lastly, addressing the performance gap observed in SSA validation as opposed to the relatively larger BraTS glioma (GLI) validation set, two strategies are proposed: fine-tuning SSA cases using the GLI+SSA best-pretrained 2D fullres model at 300 epochs, and introducing a novel neural style transfer-based data augmentation technique for the SSA cases. This investigation underscores the potential of enhancing brain tumor prediction within SSA's unique healthcare landscape.
\end{abstract}
\section{Introduction}\label{sec:intro}
Brain tumors present a substantial health challenge in Africa. 
The efforts of research on brain tumors have barely made any positive change in survival rate in low-and middle-income countries (LMICs). The rate of mortalities from glioma is among the highest in the world, with the Sub-Saharan Africa experiencing a rise of 25\% \cite{adewole2023brain}.

Accurate segmentation of distinct sub-regions within gliomas such as peritumoral edema, necrotic core, enhancing, and non-enhancing tumor core, based on multimodal MRI scans, hold clinical relevance for the diagnosis, prognosis, and treatment of brain tumors.
Accurately delineating the regions of interest within a tumor provides essential insight about its size, location, and shape, enabling the determination of the extent of tumor involvement \cite{liu2014survey}.

However, the segmentation of these sub-regions presents a formidable challenge due to the heterogeneity of brain tumors\cite{feng2020brain} and in resource-constrained settings such as the SSA, due to the propensity for suboptimal image contrast and resolution\cite{adewole2023brain} from lower quality MRI scanners and lack of availability of advanced imaging techniques\cite{anazodo2023framework}. These challenges raise uncertainty about the feasibility of implementing ML methods for clinical purposes \cite{anazodo2023framework} \cite{zhang2022stroke}. Furthermore, suboptimal image contrast and resolution \cite{lin2019deep}, necessitates advanced image pre-processing to enhance their resolution before employing ML techniques for tasks like tumor segmentation, classification, or outcome prediction \cite{adewole2023brain}. 

Therefore, this study aims to develop a generalizable deep learning-based brain tumor segmentation method. Our approach addresses the challenge of lower quality MRI scans in the SSA by developing an effective model that handles data variability. Through advanced preprocessing and optimized segmentation, we aim to create a robust model that improves diagnostic outcomes in under-resourced settings.

\subsection{Related Work}
Neural style transfer (NST) has been employed as a data augmentation technique in various medical domains, including COVID-19 diagnosis classification \cite{hernandez_improving_2021} and 3D cardiovascular MRI image segmentation \cite{ma_neural_2019}. However, this study marks the first application of NST to brain tumor segmentation within a Sub-Saharan Africa context.
Moreover, Tomar et.al. \cite{tomar_self-supervised_2022} explored a self-supervised style transfer technique as data augmentation to improve brain tumor segmentation performances. This comprehensive methodology was, however, more computationally expensive than the NST approach. Bouter et al. \cite{de2022deep} demonstrated the feasibility of artificially creating super-resolution MRI images from low-resolution counterparts, indicating that such an approach could be leveraged for the BRaTS-2021 dataset. Lastly, the work conducted by Sendra et al. \cite{sendra2023generalisability} addressed similar domain-shift challenges using comparable approaches. Their findings suggested that employing transfer learning for domain adaptation could integrate modest-sized African samples into extensive databases of developed nations. Notably, both studies highlighted the imbalance between African and high-resource country cases, with Sendra et al. studying 25 patients from five African centers, while our study examined 60 African cases.

\section{Methods}
\label{sec:Methods}

\subsection*{Implementation}
Our solution, implemented with PyTorch, is an extension of nnUNet \cite{isensee2021nnunet} which is publicly available on GitHub: \url{https://github.com/MIC-DKFZ/nnUNet}. The baseline model training and inference were done with mixed precision to minimize costs (i.e. time and memory). The experiments were run on Tesla T4 Turing GPUs and NVIDIA’s V100 system on Compute Canada cluster. We then stored both the latest and best checkpoint models based on the Dice score on the validation dataset for use during inference.

\subsection*{Datasets}
This work was conducted as part of the BraTS 2023 Challenge, where the datasets pre-selected by the organizers were used. The datasets were multi-center MRI scans of 1251 adult glioma (GLI) cases from the 2021 Continuous Evaluation sub-challenge\cite{BraTS2021} and 60 adult glioma cases acquired in SSA (SSA) from the BraTS-Africa sub challenge\cite{adewole2023brain} - the largest publicly available African adult glioma MRI data at the time of the 2023 challenge. Thus, a total of 1311 MRI scans of adults with pre-operative glioma including both low-grade glioma (LGG) and high-grade glioma (GBM/HGG) were used to train and validate the proposed model. The MRI scans comprised of routine T1-weighted (T1), post-contrast T1-weighted (T1ce), T2-weighted (T2) and T2-weighted Fluid Attenuated Inversion Recovery (FLAIR) images acquired as part of standard of care\cite{BraTS2021} \cite{adewole2023brain}. Each case also contained pre-labelled brain tumor sub-region masks, namely, necrotic tumor core (NCR), enhancing tumor (ET), and peritumoral edematous tissue (ED). 

The datasets were split into GLI (1251 cases) and GLI+SSA (1311) and used separately to train the model, while a five-fold cross-validation approach was used to evaluate the performance of the model. Four cases from the SSA dataset were excluded as outliers based on image quality inconsistencies. To investigate the impact of the outliers in real-world applications, the GLI+SSA data were also trained excluding the four outlier cases (GLI+SSA2; 1307 cases).

\subsection*{Data Preprocessing}
The data were preprocessed following Futrega et.al's Optimized U-Net approach \cite{futrega2021optimized}, which involved foreground cropping operation, intensity normalization, and resampling. These preprocessing steps were taken to establish data coherence among the multi-center data, enhance image quality, and standardize the format for subsequent processing steps. Commencing with loading the dataset, organized in alignment with a designated data path, we extracted crucial metadata from a JSON file. 
Enhancing image quality and uniformity was realized through the application of a crop foreground operation. This step effectively removed extraneous background regions.
Additionally, the process encompassed intensity normalization. Notably, for MRI scans, normalization was confined to non-zero regions. Addressing variations in voxel spacing among diverse scans, resampling played a pivotal role. The aforementioned measures are crucial in reducing variability in input data and improving reliability of the model's predictions\cite{futrega2021optimized}.


\subsection{Baseline Model (Optimized U-Net)}
The Optimized U-Net model \cite{futrega2021optimized} was used as our baseline model, with deep supervision to improve the gradient flow by calculating the loss function at various decoder levels. Here, each experiment was trained for 2, 5, 10, and 30 epochs using the Adam optimizer with varying learning rates for the three experiments: 1) GLI, 2) GLI+SSA, and 3) GLI+SSA2. 

\subsection{nnU-Net Model}
nnU-Net, an image segmentation method introduced in \cite{isensee2021nnunet}, adapts to specific datasets by autonomously configuring a U-Net-based segmentation pipeline. It simplifies model training by creating multiple U-Net configurations for different datasets, effectively handling diverse input modalities and class imbalances. Notably, nnU-Net version 2 was used given its ser-friendly development framework.

\subsubsection{Two- and three-dimensional (2D \& 3D) Configuration}

We have employed a 3D full resolution with a batch size of 2, a patch size of [128, 128, 128], 32 U-Net base features, a per-stage encoder and decoder of [2, 2, 2, 2, 2, 2], kernel sizes of [[1, 1, 1], [2, 2, 2], [2, 2, 2], [2, 2, 2], [2, 2, 2], [2, 2, 2]], and convolution kernel sizes of [[3, 3, 3], [3, 3, 3], [3, 3, 3], [3, 3, 3], [3, 3, 3], [3, 3, 3]] on both the GLI dataset and the GLI+SSA dataset. We also employed a 2D full resolution with a batch size of 105, a patch size of [192, 160], 32 U-Net base features, a per-stage encoder and decoder of [2, 2, 2, 2, 2, 2], kernel sizes of [[1, 1], [2, 2], [2, 2], [2, 2], [2, 2], [2, 2]], and convolution kernel sizes of [[3, 3], [3, 3], [3, 3], [3, 3], [3, 3], [3, 3]] on the GLI and GLI+SSA dataset. Each experiment was trained for 2, 5, 10, 30, and 300 epochs using the Adam optimizer with varying learning rates. 

\subsection{Proposed Methods: Neural Style Transfer Augmentation and 2D Full-Res Model Finetuning}
In the context of resource-constrained settings where limited training data continues to pose a challenge, we employed neural style transfer (NST), first proposed by Gatys et.al. \cite{gatys_neural_2015}, as a data augmentation technique to enhance the effectiveness of our model training process. 
NST is a technique rooted in deep learning that enables the separation and combination of content and style aspects within images. 
Specifically, our application leveraged NST to enhance the SSA MRI image quality as a data augmentation method. This technique involved using an SSA MRI image as the content (or source) image and randomly pairing with a GLI MRI image as the style (or target) image. The NST process entailed adapting the stylistic features of the GLI image onto the SSA image, thereby creating new, augmented training samples for SSA cases.
Using the Keras functional API, the intermediate layers of a pretrained VGG19 image classification network\cite{simonyan2015vgg} were used as a feature extractor to obtain the content and style representations of an image. The neural style transfer algorithm computes a loss by evaluating the discrepancies between the stylistic and content features of the generated image and target images. The style loss measures stylistic differences using Gram matrices of feature maps, while the content loss quantifies content dissimilarity through feature map comparisons.

\subsubsection{Overall workflow}
The proposed methodology is illustrated below in Figure \ref{fig:proposed methodology workflow}. At the time of submission for this short paper, we hereby present results for the nnU-Net best 2D and 3D models, as well as the best pre-trained 2D fullres nnU-Net model trained from GLI+SSA training data fine-tuned on the original and NST-augmented SSA training data. The 5-fold cross-validation model evaluation was based on mean Dice Similarity Coefficient (DSC) after each epoch. A Paired Samples t-test was conducted to compare model performances between datasets and as a function of training time/epochs (significance at p<0.05). Our code was made available on Github: \url{https://github.com/CAMERA-MRI/SPARK2023/tree/main/SPARK_BTS_KIFARU}.

\begin{figure}[!ht]
    \centering
    \resizebox{0.8\textwidth}{!}{\input{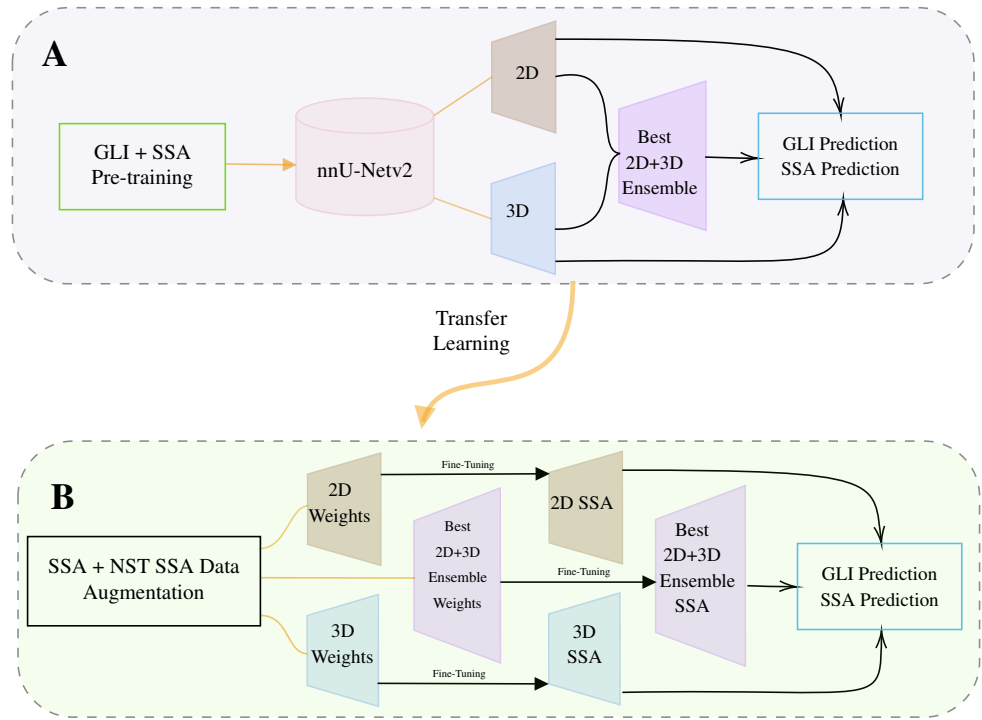}}
    \caption{Fine tuning small-scale SSA data with neural style transfer data augmentation techniques (B) with pretrained weights from large-scale GLI data via transfer learning (A).}
    \label{fig:proposed methodology workflow}
\end{figure}

\section{Results} \label{sec:results}

\subsection{Baseline Model (Optimized U-Net)}
To establish a robust evaluation, we conducted experiments using the Optimized U-Net \cite{futrega2021optimized} on both the GLI and GLI+SSA datasets as part of our foundational benchmarking process. This comparison aimed to elucidate how distributional shifts within training data can influence cross-validated model performance. Table \ref{tab1:Optimized U-Net Results for all datasets} presents outcomes from both datasets. It includes their Dice Similarity Coefficient(DSC), training, and validation losses for the GLI and GLI+SSA datasets, with or without four outlier cases (00051, 00097, 00041, and 00084), as shown in Figure \ref{Fig: mask SSA} in Appendix. We halted training at 30 epochs for comparative purposes with an experiment using the complete dataset for the same epoch count. From the aforementioned results, a decrease in model performance was observed from a Dice score of 0.89 for the GLI trained dataset to 0.88 for the GLI+SSA dataset at 30 epochs (Table 1). There was no statistical significant difference between the Dice scores of the GLI and GLI+SSA models (p>0.05) or between models when outliers were excluded (p>0.05) (Table 1). However, this exclusion was specific to this experiment, aimed at probing the domain shift issue in multi-institutional data, particularly in the SSA context.

\begin{table}[htbp]
  \centering
  \caption{Comparison of the GLI, GLI+SSA, and GLI+SSA2 (Excl. 4 SSA outliers) Datasets for the Optimized U-Net.}
  \makebox[\textwidth]{ 
  \begin{tabular}{@{}c|ccc|c@{}}
    \toprule
    \multirow{2}{*}{\textbf{Dataset}} & \multicolumn{3}{c|}{\textbf{Results}} & \multirow{2}{*}{\textbf{Paired t-test}} \\
    \cmidrule{2-4}
    & \textbf{Dice Similarity Coefficient} & \textbf{Train Loss} & \textbf{Val Loss} & \\
    \midrule
    \multirow{4}{*}{GLI} & 0.79 & 1.83 & 0.21 & \multirow{4}{*}{GLI vs. GLI+SSA:} \\
    & 0.84 & 1.12 & 0.15 & \multirow{4}{*}{\textbf{t-stat=1.82, p=0.17, df=3}} \\
    & 0.86 & 0.85 & 0.14 & \\
    & \textbf{0.89} & 0.59 & 0.10 & \\
    \midrule
    \multirow{4}{*}{GLI+SSA} & 0.77 & 1.90 & 0.23 & \multirow{4}{*}{GLI+SSA vs. GLI+SSA2:} \\
    & 0.84 & 1.01 & 0.15 & \multirow{4}{*}{\textbf{t-stat=-0.51, p=0.65, df=3}} \\
    & 0.86 & 0.76 & 0.14 & \\
    & \textbf{0.88} & 0.57 & 0.12 & \\
    \midrule
    \multirow{4}{*}{GLI+SSA2} & 0.79 & 1.58 & 0.20 & \\
    & 0.83 & 1.01 & 0.17 & \\
    & 0.86 & 0.56 & 0.14 & \\
    & \textbf{0.88} & 0.55 & 0.11 & \\
    \bottomrule
  \end{tabular}
  }
  \label{tab1:Optimized U-Net Results for all datasets}
  \vspace{0.5cm} 
\end{table}

\subsection{nnU-Net Model (version 2)}

We compared the 3D fullres versus 2D fullres configurations of nnU-Net trained on the GLI+SSA dataset. The best model was considered to have obtained the best 5-fold cross-validation averaged Dice Similarity Coefficient.

\subsubsection{Configuration name: 3D full resolution}
 Here, we employed the 3D configuration models for the GLI and GLI+SSA datasets, as represented in Table \ref{Tab2: 3D full res nnU-Net 2021 vs. GLI + SSA}. It is evident that the 300 epoch model achieved the best pseudo Dice score at 0.95 and 0.93 for the GLI and GLI+SSA datasets respectively; followed by the 30 epoch model at 0.90 for both the datasets. The paired t-test comparisons between models, showed no statistically significant difference (p>0.05) between the pseudo Dice scores (Table 2).  

\begin{table}[!htt]
\centering
\caption{3D fullres nnU-Net configuration: A representation of the average of Pseudo Dice Score per epoch number for the GLI and the GLI+SSA datasets, with Paired t-test results included.}
\makebox[\textwidth]{ 
\begin{tabular}{cccccccc}
\toprule
\multirow{2}{*}{\textbf{Dataset}} & \textbf{Epoch} & \textbf{Learning Rate} & \textbf{Train Loss} & \textbf{Val Loss} & \textbf{Pseudo Dice} & \textbf{Epoch Time (s)} & \textbf{Paired t-test} \\
& & & & & & & \\
\midrule
\multirow{5}{*}{GLI} 
& 2 & 0.00536 & -0.67 & -0.71 & 0.78 & 169.28 & \multirow{5}{*}{GLI vs GLI+SSA} \\
& 5 & 0.00235 & -0.86 & -0.87 & 0.86 & 165.65 & \multirow{5}{*}{Pseudo Dice:} \\
& 10 & 0.00126 & -0.80 & -0.78 & 0.83 & 195.28 &  \\
& 30 & 0.00047 & -0.83 & -0.85 & \textbf{0.90} & 190.9 & \\
& 300 & 6e-05 & -0.88 & -0.87 & \textbf{0.95} & 201.65 & \multirow{5}{*}{\textbf{t-stat: 1.12}}\\
\midrule
\multirow{5}{*}{GLI+SSA} 
& 2 & 0.00536 & -0.42 & -0.64 & 0.78 & 449.16 & \multirow{5}{*}{\textbf{p-value: 0.33}}\\
& 5 & 0.00235 & -0.72 & -0.72 & 0.79 & 467.69 & \multirow{5}{*}{\textbf{df:4}}\\
& 10 & 0.00126 & -0.78 & -0.79 & 0.84 & 427.12 & \\
& 30 & 0.00047 & -0.84 & -0.85 & \textbf{0.90} & 427.3 & \\
& 300 & 6e-05 & -0.89 & -0.86 & \textbf{0.93} & 496.42 & \\
\bottomrule
\end{tabular}
}
\label{Tab2: 3D full res nnU-Net 2021 vs. GLI + SSA}
\vspace{0.5cm} 
\end{table}

\subsubsection{Configuration name: 2D full resolution }
Table \ref{Tab3: 2D full res nnU-Net 2021 vs. GLI + SSA} reveals that the 2D full-resolution nnU-Net model, trained for 300 epochs, achieved five-fold cross-validation scores of 0.91 and 0.93 for the GLI and GLI+SSA training datasets respectively. Despite the pseudo Dice score of the GLI+SSA dataset being slightly higher than that of the GLI model, the paired t-test revealed that the difference between the pseudo Dice scores is not statistically significant (p>0.05) (Table 3). Also, at 300 epochs of training, for the GLI+SSA dataset, the five-fold cross-validation performance of 2D fullres nnU-Net is equal to that of 3D (0.93). However, the GLI+SSA dataset has significantly longer training times than that of the GLI dataset, with the 3D fullres nnU-Net training faster. The leision-wise Dice Similarity Coefficients for each fold of the 2D fullres nnU-Net model are presented in Table \ref{Tab4: best 2D full res nnU-Net model from five-fold cross validatoin} of the Appendix.

\begin{table}[!ht]
\centering
\caption{2D fullres nnU-Net: A representation of the average of prediction Pseudo Dice Scores per epoch number for the GLI and the GLI+SSA datasets, with t-test results.}
\label{Tab3: 2D full res nnU-Net 2021 vs. GLI + SSA}
\makebox[\textwidth]{ 
\resizebox{\textwidth}{!}{ 
\begin{tabular}{cccccccc}
\toprule
\multirow{2}{*}{\textbf{Dataset}} & \textbf{Epoch} & \textbf{Learning Rate} & \textbf{Train Loss} & \textbf{Val Loss} & \textbf{Pseudo Dice} & \textbf{Time (s)} & \textbf{Paired t-test} \\
& & & & & & & \\
\midrule
\multirow{5}{*}{GLI} 
& 2 & 0.00536 & -0.79 & -0.82 & 0.85 & 193.98 & \multirow{5}{*}{GLI vs GLI+SSA} \\
& 5 & 0.00235 & -0.86 & -0.87 & 0.85 & 190.19 & \multirow{5}{*}{Pseudo Dice:} \\
& 10 & 0.00126 & -0.89 & -0.89 & 0.88 & 192.5 &  \\
& 30 & 0.00047 & -0.83 & -0.83 & 0.89 & 294.5 & \\
& 300 & 6e-05 & -0.89 & -0.90 & \textbf{0.91} & 330.0 & \multirow{5}{*}{\textbf{t-stat: -1.43}}\\
\midrule
\multirow{5}{*}{GLI+SSA} 
& 2 & 0.00536 & -0.80 & -0.82 & 0.84 & 337.17 & \multirow{5}{*}{\textbf{p-value: 0.22}}\\
& 5 & 0.00235 & -0.86 & -0.87 & 0.86 & 488.99 & \multirow{5}{*}{\textbf{df: 4}}\\
& 10 & 0.00126 & -0.89 & -0.88 & 0.88 & 300.19 & \\
& 30 & 0.00047 & -0.92 & -0.91 & 0.89 & 230.32 & \\
& 300 & 6e-05 & -0.91 & -0.90 & \textbf{0.93} & 501.19 & \\
\bottomrule
\end{tabular}
}
}
\vspace{0.5cm} 
\end{table}

\subsection{Fine-tuning and neural style transfer on SSA training data as a data augmentation technique together improves SSA validation results}

Despite the 2D fullres nnU-Net trained using GLI + SSA data displaying good performance in five-fold cross-validation and generating satisfactory predictions for previously unseen GLI validation data as illustrated in Figure \ref{GLI validation prediction masks} in the Appendix, its performance on SSA validation data was weaker. It particularly performed badly on the SSA MRI images that appeared to be incomplete. For example in the left image of Figure. \ref{fig:SSA_Validaion before and after finetuning from the GLI 300 epochs pretraining} in the Appendix, an empty mask was predicted. 
Our proposed NST method was used as a data augmentation method to help solve this problem. It was observed that some original low-quality, low-resolution, "incomplete" SSA MRI images were in-painted or contextualized once paired with their high-quality GLI MRI image counterpart. Examples of such random pairing visualization are shown in Appendix Figure \ref{Fig: Neural Style Transfer Data augmentation}. 

Furthermore, the best 2D fullres nnU-Net trained model at 300 epochs was used as a pretrained model to fine-tune the original 60 cases of SSA training data as well as its stylized augmented SSA data. This directly resulted in an improvement in prediction on the same unseen SSA validation example shown in the right image of Figure \ref{fig:SSA_Validaion before and after finetuning from the GLI 300 epochs pretraining} of the Appendix whereas originally an empty mask was predicted before the NST data augmentation and fine-tuning step.

\section{Discussion}

In this study, we explored the potential to enhance brain tumor segmentation in the resource-limited context of SSA by fine-tuning a pre-trained nnU-Net model with SSA training data augmented using NST, where each sample was paired with a high-quality MRI image from the BRaTS-2021 challenge dataset. \textbf{Firstly}, we highlighted the similarity in the model performance with or without the SSA datasets. The results of the Paired Samples t-test between datasets revealed that there was no significant statistical difference between the performance of the three models as shown in Table \ref{tab1:Optimized U-Net Results for all datasets}. The results showed that the model performs equally on both the GLI and GLI+SSA datasets, demonstrating the its generalizability. \textbf{\textit{Secondly}}, we conducted a comparative analysis between the performance of 3D and 2D full-resolution models, utilizing of nnU-Net (version 2) \cite{isensee2021nnunet}. Our investigation revealed that both the 2D and 3D full-res nnU-Net models trained for 300 epochs yielded an average pseudo Dice score of 0.93 for the GLI + SSA training data via a 5-fold cross-validation strategy. \textbf{\textit{Thirdly}}, our validation process on the provided GLI and SSA cases revealed a significant performance discrepancy between the GLI and SSA validation sets. This can be visually inspected by the relatively good prediction for the GLI validation data shown in Figure \ref{GLI validation prediction masks} as compared to Figure \ref{fig:SSA_Validaion before and after finetuning from the GLI 300 epochs pretraining}A in the Appendix. The fusion of the NST data augmentation with subsequent fine-tuning targeted specifically at SSA cases demonstrated significant improvements in results for the SSA validation set (Figure \ref{fig:SSA_Validaion before and after finetuning from the GLI 300 epochs pretraining}B).

An important limitation of our study was the scarcity of African datasets. Future work should extend this approach to a larger African dataset to enhance its applicability. Thus, as proposed in the overall methodology workflow in Figure \ref{fig:proposed methodology workflow}, one may ensemble the best 2D fullres and 3D full res nnU-Net model trained from the combined GLI and SSA training data, before repeating the fine-tuning experiment with the NST data augmented SSA in addition to the original SSA training data. Moreover, the availability of higher-quality GLI data should be exploited such that more NST random pairing with the limited SSA training data could be experimented.

\section{Conclusion}\label{sec:conclusion}
As a part of the BRaTS-Africa 2023 Challenge, we have established the viability of enhancing brain tumor prediction within the limited-resource context of Sub-Saharan Africa (SSA). By utilizing a pre-trained and high-performing 2D fullres nnU-Net model, we achieved refinement through fine-tuning using SSA training data augmented via neural style transfer. This methodology underscores the potential for notable performance improvements within SSA's unique healthcare setting. 

\section*{Acknowledgments}
 The authors would like to thank the following instructors of the Sprint AI Training for African Medical Imaging Knowledge Translation (SPARK) Academy 2023 summer school on deep learning in medical imaging for providing insightful background knowledge on brain tumors that informed the research presented here; Craig Jones, Eranga Ukwatta, Esin Ozturk-Isik, Evan Calabrese, Jeff Rudie, Konstantinos Gousias, MacLean Nasrallah, Malhar Patel, Mueez Waqar, Nicole Levy, Peizhi Yan, Piotr Pater, Ujjwal Baid, and Yihao Liu. The authors would also like to thank Talha Chaudhry for his clinical input with regard to the Sub-saharan Africa dataset and Linshan Liu for administrative assistance in supporting the SPARK Academy training and capacity-building activities. The authors acknowledge the computational infrastructure support from the Digital Research Alliance of Canada (The Alliance) and knowledge translation support from the McGill University Doctoral Internship program through student exchange program for the SPARK Academy. The authors are grateful to McMedHacks for providing foundational information on python programming for medical image analysis as part of the 2023 SPARK Academy program. This research was funded by the Lacuna Fund for Health and Equity (PI: Udunna Anazodo, grant number 0508-S-001) and National Science and Engineering Research Council of Canada (NSERC) Discovery Launch Supplement (PI: Udunna Anazodo, grant number DGECR-2022-00136).

 \section*{Author Contributions}
RC \& JS: conceptualization, method design, software, statistical analysis, writing and editing; RC, DZ, \& UCA: funding and technical support, data curation, reviewing and editing; TC: reviewing of MRI images; KM: resources support; YZ: supervision, conceptualization, method design, software, statistical analysis, writing, reviewing and editing.

\bibliographystyle{unsrt}  
\bibliography{references}


\newpage
\renewcommand\appendixname{Appendix} 
\appendix
\renewcommand\thesection{\Alph{section}}
\setcounter{page}{1}
\setcounter{figure}{0}
\setcounter{table}{0}

\section{Appendix}

\begin{figure}[!ht]
  \centering
    \includegraphics[width=0.8\textwidth, angle=0]{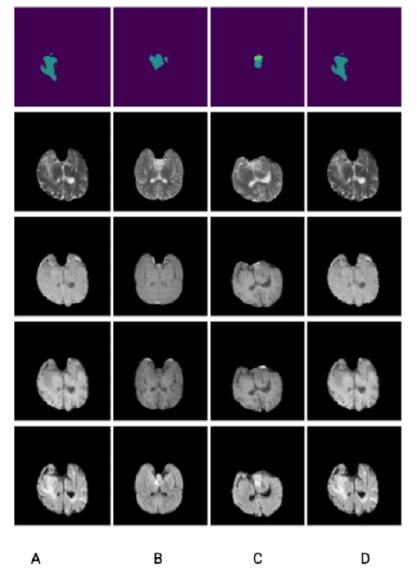}
  \caption{Excluded SSA training Cases A) 00051, B) 00097, C) 00041, \& D) 00084 only for the optimized UNet baseline experiment [Predicted masks (top row), T2 (second row), T1ce (third row),T1 (fourth row), and T2-FLAIR (bottom row)]. This was not employed for the rest of the experiments. The predicted masks are shown with  color coding as follows: background: purple, necrotic tumor core (NCR): blue, enhancing tumor (ET): yellow, peritumoral edematous tissue (ED): turquoise.}
  \label{Fig: mask SSA}
\end{figure}

\begin{figure}[!ht]
  \centering
    \includegraphics[width=0.8\textwidth]{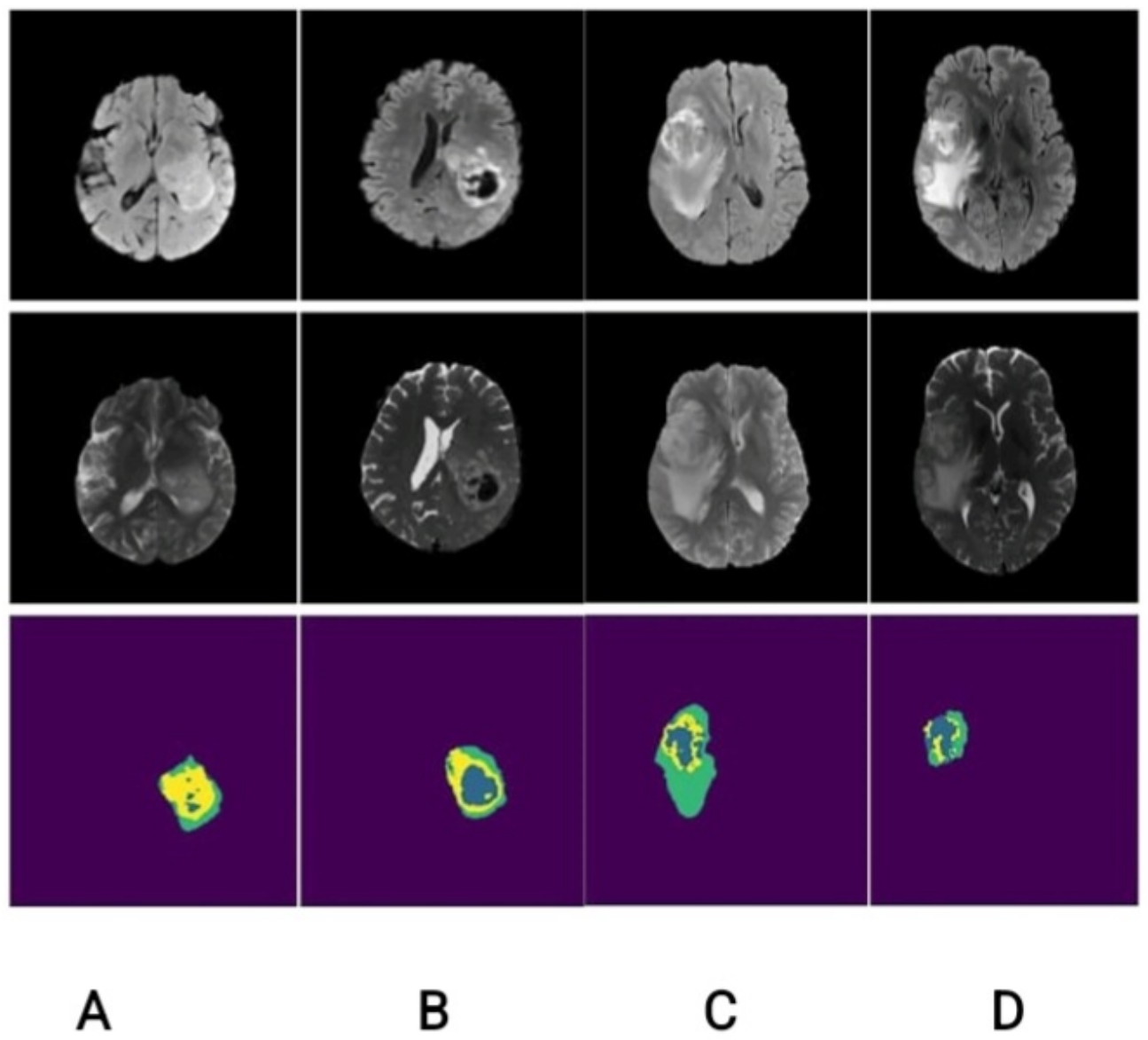}
  \caption{Predicted Masks for validation data BraTS-GLI-00001-000 (A), BraTS-GLI-00001-001 (B), BraTS-GLI-00013-000 (C), and BraTS-GLI-00013-001 (D), [T1 (top row), T2 (bottom row)] cases using the well-performing best 2D fullres nnUNet model without fine-tuning. The predicted masks (bottom row) are shown with color coding as follows: background: purple, necrotic tumor core (NCR): blue, enhancing tumor (ET): yellow, peritumoral edematous tissue (ED): turquoise.}
  \label{GLI validation prediction masks}
\end{figure}

\begin{figure}
    \centering
    \includegraphics[width=1\linewidth]{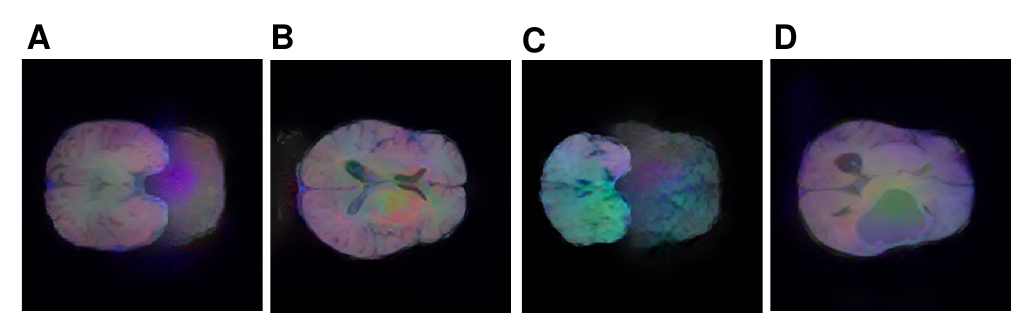}
    \caption{Examples of Neural style transfer results between the high-resolution GLI MRI images (style image) and the low-resolution SSA MRI images (content image) via one-to-one random pairing. This was used a data augmentation approach.}
    \label{Fig: Neural Style Transfer Data augmentation}
\end{figure}

\begin{figure}[!ht]
    \centering
    
    \includegraphics[width=0.5\textwidth]{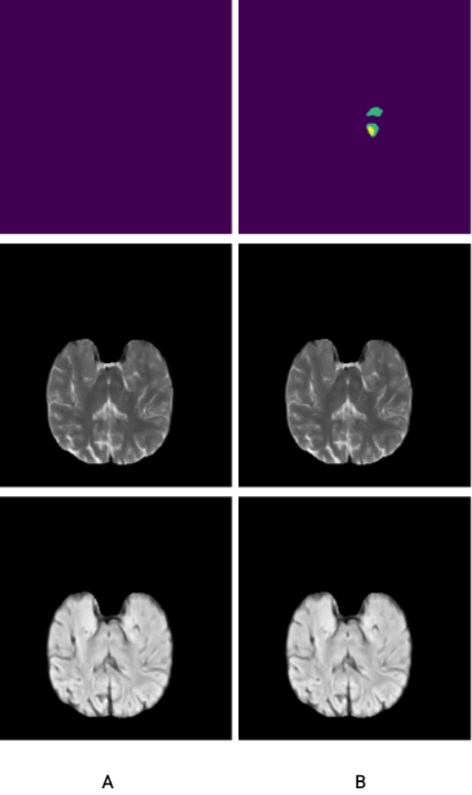}

    \caption{Tumor segmentation improvement for SSA validation case BraTS-SSA-00192-000 before (A) and after (B) neural style transfer data augmentation. [T1 (bottom row), T2(middle row)]. This was also after fine-tuning on SSA training data only from the best GLI pretrained model at the best 2D fullres nn-Unet. The color coding is the following: background: purple, necrotic tumor core (NCR): blue, enhancing tumor (ET): yellow, peritumoral edematous tissue (ED): turquoise.}
    \label{fig:SSA_Validaion before and after finetuning from the GLI 300 epochs pretraining}
\end{figure}

\begin{table}[htbp]
  \centering
  \caption{Lesion-wise dice score from five-fold cross-validation for the best 2D fullres nnU-Net trained model.}
  \label{Tab4: best 2D full res nnU-Net model from five-fold cross validatoin}
  \begin{tabular}{@{}c|ccc|c@{}}
    \toprule
    \multirow{2}{*}{\textbf{Model}} & \multicolumn{3}{c|}{\textbf{Dice Score}} & \multirow{2}{*}{\textbf{Epochs}} \\
    \cmidrule{2-4}
    & \textbf{Dice\_ET} & \textbf{Dice\_TC} & \textbf{Dice\_WT} & \\
    \midrule
    \multirow{4}{*}{Fold 0} 
    & 0.8689 & 0.8205 & 0.8082 & 2 \\ 
    & 0.9131 & 0.8534 & 0.8304 & 5 \\
    & 0.8963 & 0.8231 & 0.8415 & 10 \\
    & 0.9388 & 0.9031 & 0.8991 & 30 \\
    & \textbf{0.9471} & \textbf{0.9179} & \textbf{0.9179} & 300 \\
    \midrule
    \multirow{4}{*}{Fold 1} 
    & 0.8745 & 0.7932 & 0.776 & 2 \\
    & 0.9234 & 0.866 & 0.8509 & 5 \\
    & 0.9327 & 0.8839 & 0.8922 & 10 \\
    & 0.9294 & 0.8949 & 0.8958 & 30 \\
    & \textbf{0.9488} & \textbf{0.9051} & \textbf{0.8890} & 300 \\
    \midrule
    \multirow{4}{*}{Fold 2} 
    & 0.8805 & 0.8024 & 0.8212 & 2 \\
    & 0.9178 & 0.8568 & 0.8352 & 5 \\
    & 0.9388 & 0.9062 & 0.8977 & 10 \\
    & 0.9469 & 0.9145 & 0.902 & 30 \\
    & \textbf{0.9369} & \textbf{0.8955} & \textbf{0.8989} & 300 \\
    \midrule
    \multirow{4}{*}{Fold 3} 
    & 0.8197 & 0.7734 & 0.7608 & 2 \\
    & 0.9129 & 0.837 & 0.8247  & 5 \\
    & 0.9303 & 0.8709 & 0.8414 & 10 \\
    & 0.9395 & 0.9096 & 0.9009 & 30 \\
    & \textbf{0.9441} & \textbf{0.9101} & \textbf{0.9011} & 300 \\
    \midrule
    \multirow{4}{*}{Fold 4} 
    & 0.8706 & 0.8028 & 0.7922 & 2 \\
    & 0.8914 & 0.8247 & 0.8004 & 5 \\
    & 0.9248 & 0.8637 & 0.8592 & 10 \\
    & 0.9294 & 0.8949 & 0.8958 & 30 \\
    & \textbf{0.9459} & \textbf{0.9222} & \textbf{0.9211} & 300 \\
    \midrule
    \bottomrule
  \end{tabular}
\end{table}

\end{document}